# Enhanced Prolog Remote Predicate Call Protocol

Alin Suciu, Kalman Pusztai, Andrei Diaconu
*Technical University of Cluj-Napoca*
*Department of Computer Science*
*{Alin.Suciu, Kalman.Pusztai}@cs.utcluj.ro, andi@bdumitriu.ro*

**Abstract**

*Following the ideas of the Remote Procedure Call model, we have developed a logic programming counterpart, naturally called Prolog Remote Predicate Call (Prolog RPC) [1]. The Prolog RPC protocol facilitates the integration of Prolog code in multi-language applications as well as the development of distributed intelligent applications. One use of the protocol's most important uses could be the development of distributed applications that use Prolog at least partially to achieve their goals. Most notably the Distributed Artificial Intelligence (DAI) applications that are suitable for logic programming can profit from the use of the protocol. After proving its usefulness, we went further, developing a new version of the protocol, making it more reliable and extending its functionality. Because it has a new syntax and the new set of commands, we call this version Enhanced Prolog Remote Procedure Call. This paper describes the new features and modifications this second version introduced. One difference is that a connection comprises two modes, clients being able to switch between them by logging on, respectively logging off. New features include capturing of Prolog program output, and modifying Prolog machine flags. The operation of executing predicates has also been redesigned.*

## 1. Introduction

Logic Programming, and the language Prolog in particular, has been proven to be very useful for implementing applications from various fields of Artificial Intelligence (e.g. Expert Systems, Machine Learning, Search, Reasoning, Planning, Natural Language Processing, Deductive Databases, Data Mining, etc) [3], [8]. Its declarative nature and high level programming features makes it also very suitable for Software Engineering, e.g. for rapid prototyping and programming-in-the-large.

Following the success of Sun's Remote Procedure Call (RPC) model [10], [2], various attempts were made do adapt the ideas of RPC to the logic programming paradigm [4], [9], and Prolog in particular [5], [7], [13]. The logic programming counterpart of the Remote Procedure Call must naturally be the Remote Predicate Call [1]. Based on the past version, we developed a new version, called Enhanced Prolog RPC making it more reliable and extending its functionality.

The paper presents the new features and main modification this new version has brought, with its new syntax, and extended set of commands.

After presenting the fundamentals of the Enhanced Prolog RPC protocol in Section 2, a detailed description of the Enhanced Prolog RPC follows in Section 3. In Section 4 we present some possible applications and Section 5 deals with some implementation issues. Finally in Section 6 some conclusions are drawn and some further developments are mentioned.

## 2. Fundamentals of the Enhanced Prolog RPC protocol

The main objective of the Prolog RPC protocol is similar to the classical RPC objective: to be able to call Prolog predicates (i.e. ask queries) remotely. We can say that the Prolog RPC protocol is a:
1. connection-oriented protocol
2. client-server protocol
3. request-response protocol
4. platform independent protocol
5. language independent protocol

It is a client-server protocol which operates based on the following assumptions:
1. the server must contain a Prolog engine that enables it to execute Prolog queries; the server can be written entirely in Prolog
2. the client can call Prolog predicates (ask Prolog queries) that are stored on the server, or may be uploaded by the client, and receives the answers in a standard manner; the client can be written in any programming

language but it must correctly implement the Prolog RPC protocol

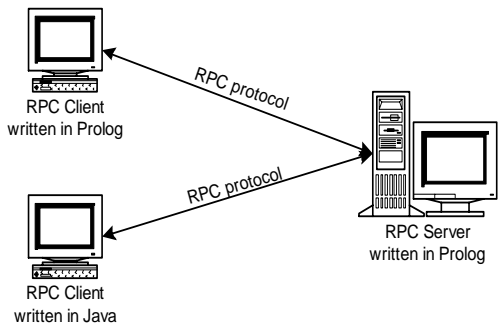

**Figure 1. The Prolog RPC protocol**

The following generic algorithm describes the operation of the protocol, under the ideal assumption of the absence of any connectivity errors:

Step 1. The client establishes a connection with the Prolog RPC server
Step 2. The client authenticates
Step 2. Bidirectional data transfer takes place
Step 3. The client closes the connection

The actual remote predicate call takes place in the third step of the algorithm which consists of a series of request-response messages between the client and the server. What is particularly interesting here is that not only predicates that are already loaded in the server may be called. The client can send Prolog code to the server which is automatically loaded into the server's database and therefore can be subsequently called by the client(s).

There are a number of security issues that must be addressed, the most important one being related to allowing the execution of potentially dangerous code sequences (e.g. operating system calls). Security policies can be designed for the server so that the security risks are minimized or even eliminated.

The protocol ensures the lose coupling between Prolog and any other programming language. Basically any computer program, written in any language, under any platform can execute Prolog code by connecting to a Prolog RPC server. Whenever a part of a problem can be solved more suitable in Prolog (e.g. Symbolic Processing, Expert Systems, Machine Learning, Search, Reasoning, Planning, Natural Language Processing, Deductive Databases, Data Mining, etc) one can just connect to a Prolog RPC server, load the code there, ask the queries, get the results needed, return and continue the execution in the original program.

## 3. The Enhanced Prolog RPC protocol

The major differences between the first version and the enhanced one are the existence of two modes a connection may have, and the simplification of the messages passed between clients and server.

An Enhanced Prolog RPC connection comprises two modes, corresponding to two states. In the first state, client opened a connection, but has not identified himself. In the second state, client has identified, supplying a username and a password. Clients may switch between these two states using the commands "login", and respectively "logout". After login, a session will be established between the client and the server, and security policies are applied. Each user has special rights according to the user rights policy adopted on the server (e.g. some of them can only list the predicates; others may execute them, etc). At any moment, either the client or the server can terminate this session, by simply closing the connection.

Before login, any client may send the following commands: "ping", "ver", "status", and "login". After login, the available commands are: "ping", "ver", "status", "list", "capture_output", "release_output", "execute", "add_file", "comment", "uncomment", "detail", "set_flag", "delete", "delete_all", and "logout".

The server will usually respond to each request with one the following keywords:
"yes" if everything was ok
"no" if the execution of the command failed
"error" if an error occurred

All the messages passed between the client and the server are Prolog terms, ending with a period and a new-line character (\n). Each message comprises only one term, unlike the first version of the protocol. This approach simplifies a lot the Prolog implementation of the RPC server. The detailed description of each message type is given below.

### 3.1. Commands available in both modes

#### 3.1.1. Version.
Request:
    ver.\n
Response:
    yes('version number').\n
The version command always succeeds, and it returns the version of the protocol. This command is available in both modes.

#### 3.1.2. Ping.
Request:
    ping.\n
Response:
    yes('pong').\n

The ping command is used to test connection or server.

### 3.1.3. Status.
Request:
   status.\n

Response:
   yes([sesionID/username, ListOfAllowedCommands]).\n
   no.

The version command returns the ID, username and the list of permitted commands, if user is logged on, or "no" otherwise.

### 3.2. Commands available before login

### 3.2.1. Login.
Request:
   login(id, login, password).\n
Response:
   yes.\n
   error('access denied').\n

The client requests to open a connection "id" using the specified login and password. The existence of a connection identifier allows the creation of multiple connections and ensures the persistence of the connection.

If the response is "yes.\n" the connection's state changes, and client may send the commands available after login.

### 3.3. Commands available after login

### 3.3.1. Execute (first solution).
Request:
   execute(query).\n
Response:
   yes([inst var list]).\n
   no.\n
   error(error message).\n

If the query succeeded, the first solution was computed and the response token "yes" is returned together with the values of each variable in the query; if the query failed, the "no" reply is returned. The error signals the fact that the user doesn't have the right to execute queries, the predicate does not exist, or if any error occurs. If succeeded, the execute command must be followed either by the "next" command. Or by the "eop" command, which signals that no more solutions are needed.

### 3.3.2. Next (solution).
Request:
   next.\n
Response:
   yes([inst var list]).\n
   no.\n
   error(error message).\n

In case additional solutions must be computed, the client can do so by issuing a "next" request until the "no" reply is returned, which means that there are no more solutions. The meanings of "yes" and "error" are the same as above.

### 3.3.3. End of predicate (Eop).
Request:
   eop.\n
Response:
   yes.\n
   error(error message).\n

Informs the server that no more solutions are needed for the current predicate.

### 3.3.4. Capture output.
Request:
   capture_output.\n
Response:
   yes.\n
   error(error message).\n

Requests for output generated by the Prolog engine to be redirected on the connection stream. Calls of predicate "write" will send the data on the connection stream.

### 3.3.5. Release output.
Request:
   release_output.\n
Response:
   yes.\n
   error(error message).\n

It is the opposite command for capture output. It disables the output capturing. Nothing happens if output has not been captured before.

### 3.3.6. Add file.
Request:
   add_file.\n Predicate List "end_of_file"
Response:
   yes.\n
   error(error message).\n

This message allows the client to upload an assert a list of predicates, usually the list being the content of a file. Once the predicates were uploaded on the server they can be executed using an "execute" message.

### 3.3.7. Comment.
Request:
   $comment(comment_text).\n
Response:
   yes.\n
   error(error message).\n

Every Prolog predicate from the server has an associated comment which usually should hold useful information about the predicate. This command assigns a comment to a predicate.

### 3.3.8. Uncomment.
Request:
    $uncomment(predicate).\n
Response:
    yes.\n
    error(error message).\n
This message removes the comment for the given predicate.

### 3.3.9. Detail.
Request:
    detail(predicate).\n
Response:
    yes.\n
    error(error message).\n
The message is used for retrieving the comment, added with the "comment" command, of the specified predicate.

### 3.3.10. Set flag.
Request:
    set_flag(flag, old_value, new_value).\n
Response:
    yes.\n
    error(error message).\n
This message is used to update the Prolog machine flag. For this command to be successfully executed, the client needs to have special permissions.

### 3.3.12. Delete.
Request:
    delete(predicate_list).\n
Response:
    yes.\n
    error(error message).\n
Deletes the specified clauses from the database.

### 3.3.13. Delete all.
Request:
    deleteall.\n
Response:
    yes.\n
    error(error message).\n
Clears all the inserted predicates from the database.

### 3.3.14. Log out.
Request:
    logout.\n
Response:
    yes.\n
    error(error message).\n
This command is used to change the connection state. After successful logout, clients may re-login, or close the connection.
Do not use colors in your paper.

## 4. Applications

The Prolog RPC protocol presented here opens a wide range of opportunities for distributed, cross platform, cross language applications. It offers a way to loosely coupling the Prolog language with any other language, thus opening a new door for inter-language communication. Since it is commonly accepted the fact that there is no "best" programming language, one can always embed Prolog into his program for the parts that are really suitable for programming in Prolog. This eliminates the potential troubles that may occur when trying to bond Prolog more tightly with other programming languages.

Another important use of the protocol could be for the development of distributed applications that use Prolog at least partially to achieve their goals. Most notably the Distributed Artificial Intelligence applications that are suitable for logic programming can profit from the use of the protocol. I would be very easy for a client to divide the amount of work to be done into many pieces, then upload them on different RPC servers (perhaps using a planner for this) and then launching a distributed execution.

Writing distributed (even mobile) agent systems also seems an appealing opportunity which can be at least partially supported by the Prolog RPC protocol introduced here.

## 5. Implementation

For the implementation of the Prolog RPC protocol a reliable connection-oriented transport protocol must be used. Our current implementation uses sockets with the TCP/IP protocol.

The current implementation is written entirely in Sicstus Prolog [6]. It supports multiple clients simultaneously, although the Prolog engine is not a multithreaded one. It implements a fairly simple security policy, displays and logs all accesses and messages.

Client implementations consist of class libraries for the Java and C++ languages, and a client program written in Prolog.

## 6. Conclusions

Following the ideas from the Remote Procedure Call execution model we developed a logic programming counterpart, naturally called Prolog Remote Predicate Call (Prolog RPC), and extended it to create the Enhanced Prolog RPC. After presenting the fundamentals of this model, the Enhanced Prolog RPC protocol was described in detail in Section 3. The Enhanced Prolog RPC protocol facilitates the integration of Prolog code in multi-language applications as well as the development of distributed intelligent applications.

It offers a way to loosely coupling the Prolog language with any other language, thus opening a new door for inter-language communication and eliminates the potential troubles that may occur when trying to bond Prolog more tightly with other programming languages.

Another important use of the protocol could be for the development of distributed applications that use Prolog at least partially to achieve their goals. Most notably the Distributed Artificial Intelligence applications that are suitable for logic programming can profit from the use of the protocol.

Writing distributed (even mobile) agent systems also seems an appealing opportunity which can be at least partially supported by the Prolog RPC protocol introduced here.

The development of a concurrent server is under development. There are two implementations, both being a mixture of Java and Prolog.

The integration of the Prolog RPC protocol in the logic and object-oriented language LOOP [11], [12] as the LOOP RMI protocol is also considered for the next version.